# Extracting respiratory signals from thoracic cone beam CT projections

**Hao Yan[1], Xiaoyu Wang[1], Wotao Yin[2], Tinsu Pan[3], Moiz Ahmad[3], Xuanqin Mou[4], Laura Cerviño[1], Xun Jia[1] and Steve B. Jiang[1]**

[1] Center for Advanced Radiotherapy Technologies and Department of Radiation Medicine and Applied Sciences, University of California San Diego, La Jolla, CA 92037-0843, USA

[2] Department of Computational and Applied Mathematics, Rice University, Houston, TX 77005-1892, USA

[3] Department of Imaging Physics, The University of Texas, MD Anderson Cancer Center, Houston, Texas 77030, USA

[4] Institute of Image Processing and Pattern Recognition, Xi'an Jiaotong University, Xi'an, Shaanxi, 710049, China

E-mails: sbjiang@ucsd.edu, xunjia@ucsd.edu

Patient respiratory signal associated with the cone beam CT (CBCT) projections is important for lung cancer radiotherapy. In contrast to monitoring an external surrogate of respiration, such signal can be extracted directly from the CBCT projections. In this paper, we propose a novel local principal component analysis (LPCA) method to extract the respiratory signal by distinguishing the respiration motion-induced content change from the gantry rotation-induced content change in the CBCT projections. The LPCA method is evaluated by comparing with three state-of-the-art projection-based methods, namely, the Amsterdam Shroud (AS) method, the intensity analysis (IA) method, and the Fourier-transform based phase analysis (FT-p) method. The clinical CBCT projection data of eight patients, acquired under various clinical scenarios, were used to investigate the performance of each method. We found that the proposed LPCA method has demonstrated the best overall performance for cases tested and thus is a promising technique for extracting respiratory signal. We also identified the applicability of each existing method.

## 1. Introduction

Respiration-induced motion is of great concern in image-guided radiotherapy (IGRT) when treating tumors in lung or upper abdomen area. To provide reliable imaging guidance for treatment, 4D-CBCT has been developed, where all the projections are first sorted into breathing phase bins and then CBCT images at different phases are reconstructed (Sonke *et al.*, 2005; Kriminski *et al.*, 2005; Li *et al.*, 2006b; Dietrich *et al.*, 2006; Purdie *et al.*, 2006; Lu *et al.*, 2007; Guckenberger *et al.*, 2008; Leng *et al.*, 2008; Sonke *et al.*, 2009; Bergner *et al.*, 2010; Jia *et al.*, 2012). Another approach is to obtain the CBCT image with greatly reduced motion artifacts at a particular breathing phase through either hardware methods (e.g., respiratory gating) (Moseley *et al.*, 2005; Sillanpaa *et al.*, 2005; Chang *et al.*, 2006; Jiang, 2006; Thompson and Hugo, 2008; Cooper *et al.*, 2012) or software approaches (correcting motion artifacts during the reconstruction or post-processing) (Schäfer *et al.*, 2004; Li *et al.*, 2006a; Isola *et al.*, 2008; Rit *et al.*, 2009; Zhang *et al.*, 2010; Lewis *et al.*, 2011). During the treatment delivery, one can perform CBCT-based marker-less tumor tracking to predict tumor positions based on a breathing signal (Cho *et al.*, 2010; Lewis *et al.*, 2010). In all of these applications, accurate extraction of the patient-specific respiratory signal plays a vital role.

Respiration signal can be obtained via external surrogate, e.g., by tracking the displacement of markers on the thoracic/abdominal surface (Ford *et al.*, 2002; Vedam *et al.*, 2003; Pan *et al.*, 2004), measuring the air flow change of patients (spirometry) (Low *et al.*, 2003; Zhang *et al.*, 2003), or sensing the pressure variation of an elastic thoracic/abdominal belt (Dietrich *et al.*, 2006; Kleshneva *et al.*, 2006). However, one may have concerns regarding the correlation between these external surrogates and the motions of the internal anatomy (Ahn *et al.*, 2004; Hoisak *et al.*, 2004; Tsunashima *et al.*, 2004; Gierga *et al.*, 2005; Yan *et al.*, 2006). While implanted radiopaque fiducial markers (Gierga *et al.*, 2003; Shirato *et al.*, 2003) can provide more reliable respiratory signal, the invasive implanting process is undesirable and the marker migration is also a problem (Kitamura *et al.*, 2002; Nelson *et al.*, 2007).

Another group of methods is to directly extract breathing signals from the CBCT projections. One example is the Amsterdam Shroud (AS) method (Zijp *et al.*, 2004; Van Herk *et al.*, 2007), where the superior-inferior (SI) motion of the internal anatomy is enhanced by converting the projections into a so-called AS image and then the breathing signal is extracted. Another method is based on the radiological path-length change with lung volume expanding/retracting (Berbeco *et al.*, 2005). Respiratory signal can be extracted by analyzing the fluctuation of the image intensity in lung projection images. This idea was firstly applied in fluoroscopic imaging (Berbeco *et al.*, 2005) and then extended to thoracic CBCT imaging (Kavanagh *et al.*, 2009), which will be referred to as intensity analysis (IA) method in this paper. Recently, Vergalasova *et al.* reported another method that analyzes the CBCT projections in Fourier domain (Vergalasova *et al.*, 2012). It is claimed that this Fourier-transform (FT) based method works better for the cases with slow gantry rotation (i.e., 4-minute 4D-CBCT) than those with fast gantry rotation (i.e., 1-minute CBCT) (Vergalasova *et al.*, 2012).

In this paper, a novel projection-based method is proposed based on local principal component analysis (LPCA). In this method, the CBCT projection data is regarded as a 2-torus in a high-dimensional space. This 2-torus reflects the coupled motion signals from both patient respiration and gantry rotation during the CBCT scan. PCA is applied to the 2-torus piece by piece, to decouple the respiratory signal from the gantry motion signal and then paste them to yield a long sequence of extracted respiratory signal for the whole scanning period. The proposed LPCA method will be compared with the state of the art methods, *i.e.*, AS, IA and FT methods using several clinical CBCT datasets considering various clinical scenarios.

## 2. Methods and Materials

### 2.1 Existing methods

To better understand our proposed method, we will first summarize the existing methods as illustrated in figure 1. In all methods, the first step is to take logarithm of the measured projection to obtain a projection image $g(t, u, v)$, where $t$ is the gantry angle and $(u, v)$ represent the coordinates in the transverse and the axial, *i.e.*, superior-inferior (SI), directions, respectively. This projection image represents the radiological line integral at each pixel, and therefore has a better contrast than the original measurement data. Also, all methods involve the use of a region-of-interest (ROI) on the detector, which covers the whole detector along the transverse direction ($[u_l, u_U]$) and only the bottom part of the detector along the SI direction ($[v_{low}, v_{high}]$), usually being the lower thorax where the anatomical motion is relatively large. In the following sub-sections, we will briefly describe these methods.

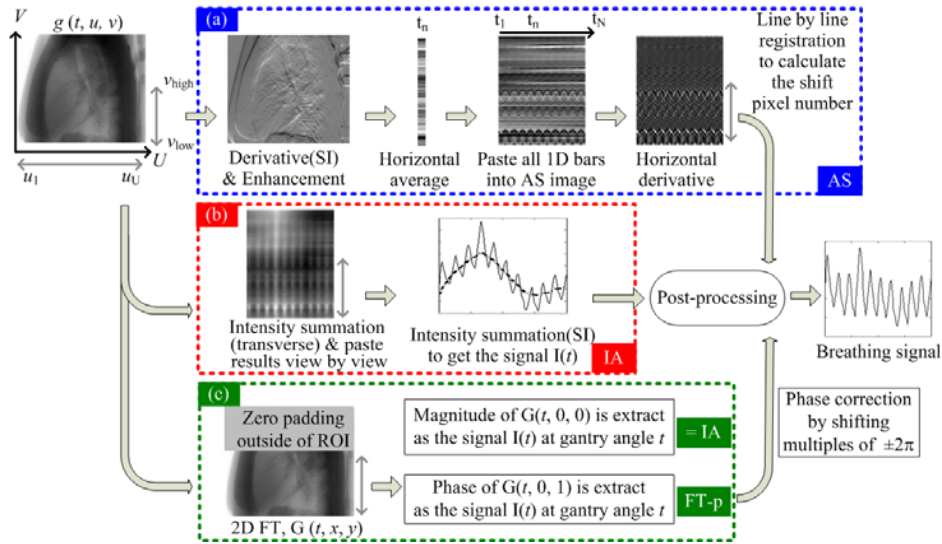

**Figure 1.** Existing projection-based breathing signal extraction methods: (a) AS method; (b) IA method; (c) FT based method. $g(t, u, v)$ represents the projection image.

### 2.1.1 AS method

The procedure of the AS method is shown in figure 1(a). Specifically, 1) the derivative of $g(t, u, v)$ is taken along the SI direction to enhance the anatomical features; 2) the resulting enhanced image is collapsed horizontally into a column by summing over all pixel intensities inside each horizontal row; 3) the columns from all projections are combined view by view to yield an AS image; 4) another derivative is performed on the AS image along the view direction to highlight the moving components; 5) a signal I($t$) is obtained by aligning each column of the results from step 4) with respect to a reference column, e.g., the first column; 6) the breathing signal is extracted from I($t$) after some post-processing, such as removing the gantry-rotation induced slow-variant component, as well as the cardiac-motion induced noise.

### 2.1.2 IA method

The intensity variations in thoracic CBCT projections are caused by two factors, specifically the patient respiration and the gantry rotation. Usually the time scale of the gantry rotation is much larger than that of the patient breathing, allowing the separation of the intensity variations caused by these two factors through applying a low-pass filter, and this is the basic idea of the IA method (Kavanagh *et al.*, 2009). Specifically, the intensity in each projection is first calculated by simply summing all pixel values in the ROI (figure 1(b)):

$$\mathrm{I}(t) = \int_{v=v_{\mathrm{low}}}^{v_{\mathrm{high}}} \int_{u=u_1}^{u_U} g(t, u, v) \, \mathrm{d}u \mathrm{d}v, \qquad t = t_1, t_2, \dots t_N. \tag{1}$$

After that, the respiratory signal can be extracted from I($t$) by applying a low pass filter to remove the slowly varying background caused by gantry rotation.

### 2.1.3 FT based method

Recently, Vergalasova *et al.* proposed two projection-based methods in Fourier domain (figure 1(c)). In their methods, zeros are padded outside the ROI to keep each projection at the original size. Then a 2-dimensional FT is performed for each projection:

$$\mathrm{G}(t, x, y) = \int_{v=v_{\mathrm{low}}}^{v_{\mathrm{high}}} \int_{u=u_1}^{u_U} g(t, u, v) \, \mathrm{e}^{-i2\pi(x \cdot u + y \cdot v)} \mathrm{d}u \mathrm{d}v, \qquad t = t_1, t_2, \dots t_N. \tag{2}$$

The first scheme uses the magnitude of G($t$, 0, 0) in (2) as the indication of the breathing signal at different view angle $t$. We find that G $(t, 0, 0) = \int_{v=v_{\mathrm{low}}}^{v_{\mathrm{high}}} \int_{u=u_1}^{u_U} g(t, u, v) \, \mathrm{d}u \mathrm{d}v$ is essentially equivalent to (1), *i.e.*, the IA method.

The second scheme is the so-called FT-phase method. Its key idea is to utilize the FT shift theorem, i.e., the space shift of a signal is reflected in a phase shift in its Fourier domain. Mathematically, denoting $p, q$ as the space shift, FT shift theorem is as follows:

$$\int_{v=v_{\mathrm{low}}}^{v_{\mathrm{high}}} \int_{u=u_1}^{u_U} g(t, u + p, v + q) \, \mathrm{e}^{-i2\pi(x \cdot u + y \cdot v)} \mathrm{d}u \mathrm{d}v$$

$$= \mathrm{G}(t. x, y) \cdot \mathrm{e}^{i2\pi(x \cdot p + y \cdot q)}, \qquad t = t_1, t_2, \dots t_N. \tag{3}$$

Vergalasova *et al.* (2012) found that the phase at (*t*, 0, 1) is more suitable for representing the breathing motion than other low frequency points. This is reasonable considering that (*t*, 0, 1) captures the basic space shift in SI direction, i.e.,$e^{i2\pi y}$, which is the dominant breathing-introduced movement in a thoracic projection image.

5          We will only investigate the FT-phase method in this paper, denoted as FT-p method, since the first scheme is equivalent to IA method.

### 2.2 Our proposed method

#### 2.2.1 The basic idea

As mentioned in section 2.1.2, the contents in the thoracic CBCT projections change due to two periodic motions: patient respiration and gantry rotation. Based on this observation, if we represent a projection with a point in a high-dimensional ambient space, these projection data points should all lie on a 2-torus. This phenomenon has been preliminarily validated by using the manifold learning technique (Wang *et al.*, 2012) .

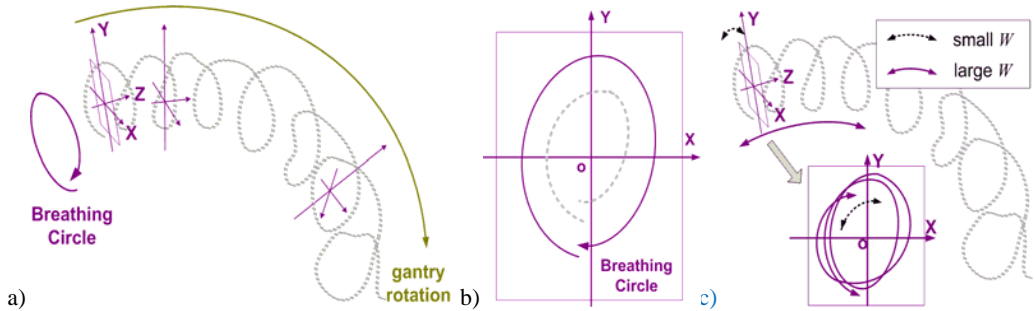

a)                                            b)                                            c)

**Figure 2.** Illustration of the basic idea of the proposed LPCA method. a) The 2-torus formed by representing the CBCT projection data as points in high dimensional space; b) A cross-section plane locally perpendicular to the length direction of the 2-torus; c) Interpreting the physical meaning of sliding window width (*W*): large *W* (purple solid line) and small *W* (black dash line)

          As indicated in figure 2(a), one direction of the torus corresponds to the breathing motion while the other to the gantry rotation. As the gantry rotates during the data acquisition, the projection data point winds around the torus along a helical trajectory, as shown in figure 2(a). If we were able to find a plane that is locally perpendicular to the direction of the torus that corresponds to gantry rotation, namely, the XoY plane as indicated in figure 2, the projection of the helical trajectory onto this plane would approximately decouple the breathing signal from the gantry rotation. Furthermore, if we pick a direction inside this plane, e.g. Y, then the projection onto this direction would give the magnitude of the breathing signal. Of course this signal is only valid for the trajectory close to this plane. However, local signals extracted at different segments of the torus can be stitched together, yielding a long sequence of global breathing signal. This is the fundamental idea of our proposed method.

#### 2.2.2 The algorithm design

To realize the idea described above, we have employed PCA, a simple and efficient technique to recognize the internal data pattern by reducing the data dimensions. Instead of the CBCT projections, the AS images are used as the input in our algorithm because it contains more distinguished breathing features, and its reduced sizes lead to the improved computational efficiency. As indicated in section 2.2.1, the 2-torus needs to be studied in a piecewise manner and each piece of 2-torus can be approximated with a linear model by performing PCA locally (LPCA). As shown in figure 3, the LPCA algorithm mainly includes two steps: extracting the foreground AS image (f-AS) and performing PCA on the f-AS image using a sliding window method.

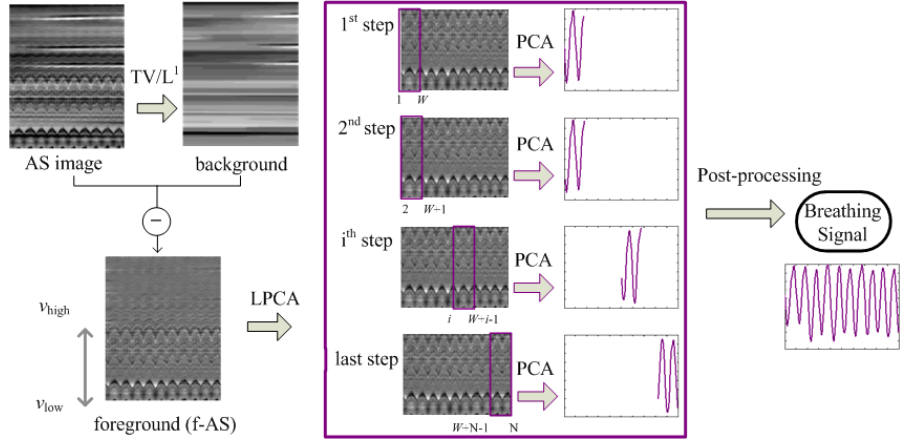

**Figure 3.** Flowchart of the proposed LPCA method. First, the background of the AS image is extracted by using a TV/L[1] method to get the foreground AS image (f-AS). Then PCA is performed on the f-AS image using a sliding window approach.

*f-AS extraction*        If we look at the original AS image, the breathing-induced wave pattern appears as a thin and periodic signal on top of a smoother background with some horizontal streaks, as shown in figure 3. Apparently, only the wavy pattern contributes to the breathing signal extraction and it is desired to separate this part as the foreground of AS image (f-AS). This operation essentially can be regarded as a further linearization to a segment of the 2-torus data. To obtain the f-AS, we employ a TV/L[1] model  (Yin *et al.*, 2006, 2007). This model extracts large-scale components $h$ from an image $f$ by solving

$$\min_h \mathrm{TV}(h) + w\|f - h\|_{\mathrm{L}^1},\tag{4}$$

where $\mathrm{TV}(h)$ is the total variation of $h$. Hence, the f-AS, which is in a small scale, can be given by $f-h$. In (Yin *et al.*, 2006), it is shown that the amount of the image components contained in $f-h$ is determined by a threshold parameter $w$. Larger $w$ value means stronger background removal, that is, if $w$ is too large, some useful anatomical features may be excluded in $f-h$; on the other hand, if the $w$ value is too small, $f-h$ will contain unwanted residual background. For the current application, $w$ value is studied and empirically selected as 0.15. We also find that a $w$ between 0.1 and 0.2 always ensures consistent LPCA performance for all the tested cases. The problem in Eq. (4) can be formulated as a maximum flow problem on a constructed flow network, which can be solved in polynomial time to the theoretically exact global solution (Goldfarb and Yin, 2009).

*Sliding window based PCA on f-AS*     PCA is performed sequentially on the f-AS image using a sliding window method with a window width of $W$ (an odd number). Assuming that the f-AS contains N columns (*i.e.*, N projections), N-$W$+1 PCAs are performed in LPCA. The detailed process is as follows:

5    1) At the step 1, PCA is performed on the first $W$ columns of the f-AS. The first principal component eigenvector (PCE) reflecting the most significant variance of data is kept as the principal direction, and the principal component coefficients (PCC) corresponding to this PCE are selected as the breathing signals for columns 1 to INT[$W$/2]+1.

10    2) At the following steps, the window slides by one column at each step and PCA is performed on the f-AS columns within the window. The correlation of each of the first five PCEs with the principal direction at the previous step is calculated. The PCE with the maximum correlation (closest to the principal direction of the previous step) is selected as the principal direction at the current step, and the corresponding PCC at the center column is selected as the breathing signal for this column (projection). As PCA cannot distinguish the positive/negative principal direction, a sign reverse may be needed to keep the current principal direction consistent with the one at the previous step.

3) At the last step, the PCCs for columns INT[$W$/2]+1 to $W$ are selected as the breathing signals.

20    We would like to discuss how this process decouples the breathing signal from the gantry rotation signal. Firstly, f-AS extracts the foreground image containing the breathing motion signal. On top of that, PCA is able to extract the most significant variance of the f-AS, corresponding to projecting the breathing signal on a plane locally perpendicular to the direction of the torus that corresponds to gantry rotation, e.g., XoY plane (figure 2). In principle, PCA finds a linear space that best approximates the data. Hence a global approximation of the entire trajectory does not work due to the essentially curved geometry. Instead, the LPCA decouples the breathing motion and gantry rotation signals by a *local* operation using a sliding window technique. In this process, $W$ represents the segment length on the 2-torus investigated. It should be small enough, such that within this range the trajectory can be approximated by a linear space to a good extent. Otherwise, many breathing cycles are involved (purple line, figure 2) and PCA find many directions, leading to a decreased extracting accuracy. On the other hand, $W$ should be larger than the number of column numbers in the f-AS corresponding one breathing cycle. Otherwise, only a portion of one breathing cycle is involved (black dash line, figure 2) and PCA cannot find the correct direction of the torus, leading to incorrect results. In the following study, we empirically select $W$ as the number of projections collected within about 1.5 breathing cycles of a typical patient. We will also present the results corresponding to different $W$ values.

40    *2.3. Evaluation*

*2.3.1 Patient data*

CBCT data of eight patients were used in this study, covering various clinical scenarios, such as different scanning protocols (*i.e.*, slow/normal scan speed, full/half fan setup, and low/normal dose level), and different image content (*i.e.*, lower/upper lung with/without the diaphragms). The first four sets of data are 4-minute 4D CBCT data. 2729, 2005, 1982 and 1679 projections were collected for patients 1-4, respectively, with rates of 5-7 frames per second, using a slow gantry speed over a 200-degree arc in a full fan mode (Lu *et al.*, 2007). Breathing signals from the commercial Real-time Position Management (RPM) system (Varian Medical System, Palo Alto, CA) are available as reference for patients 1-4, although they cannot be regarded as the ground truth. The other four sets of data (patients 5-8) are 1-minute CBCT data. 656 projections were collected over 360 degrees (10.8 frames per second) in a half-fan mode. Among these four sets of data, two were scanned with the 'thorax' mode (patients 5-6) and the other two were acquired with the 'low-dose thorax' mode (patients 7-8). The CBCT data for all eight cases were acquired using an on-board imaging (OBI) system (Varian Medical System, Palo Alto, CA). The detector contains 1024×768 pixels and the detecting region is about 40×30 cm. For normal dose data (patients 1-6), a 2×2 pixel binning was performed so that the dimension of each projection is 512×384 and the resolution is 0.784×0.784 mm. For the low dose data (patient 7-8), in order to suppress the noise, a 4×4 pixel binning was performed.

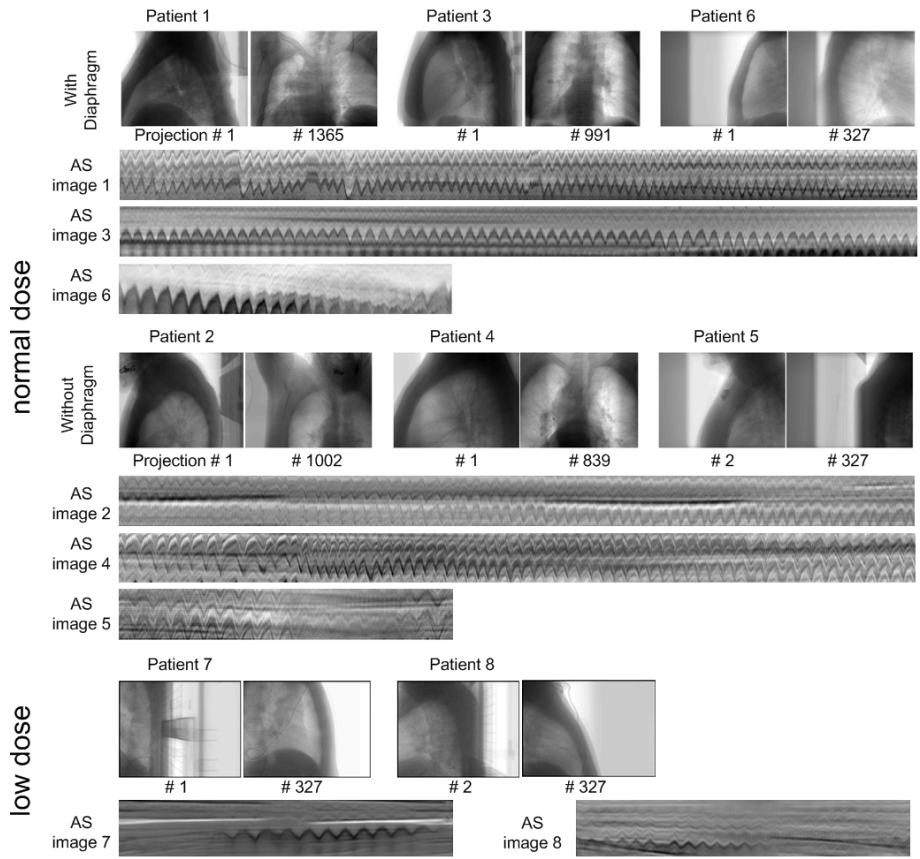

**Figure 4.** Patient datasets: typical projection images and the lower part of AS images. For patients 1-4, only ~3/4 of the columns are shown

Figure 4 shows the lateral and anterior-posterior views of each dataset, as well as the lower parts of the AS images containing distinct anatomical movement curves. The datasets are divided into three groups: 1) normal dose data with diaphragm in the projection image (patients 1, 3 and 6); 2) normal dose data without diaphragm in the projection image (patients 2, 4 and 5); and 3) low dose data (patients 7 and 8). It can be seen that the anatomical motion patterns are less distinct under half-fan mode (patients 5-8), low exposure level (patients 7 and 8), or without diaphragm in the projection image (patients 2, 4, 5).

### 2.3.2 Assessment methods

It is a difficult task to quantitatively assess the accuracy of the detected breathing signals, because strictly speaking there is no ground truth for breathing signals in clinical cases. Lung volume variation, anatomical feature (such as diaphragm) motion, as well as chest wall and abdominal surface motion are all caused by patient breathing and can be considered as breathing signals. However, these signals are not necessarily consistent with each other; phase shift may exist among them. Clinically, it is more desirable to use the motion signal from internal anatomical features as the breathing signal. In this work, we manually extract the most distinct motion patterns of the internal anatomical features from the AS image and use them as benchmark signals. But for patients 7 and 8 (low-dose cases), due to low image quality, manually extracting the benchmark signals is not feasible for some gantry angles and therefore the algorithm extracted signals are displayed with the AS images for qualitative visual inspection.

For patients 1-6, the algorithm-extracted signals are quantitatively evaluated by calculating the local correlations (Ionascu *et al.*, 2007) Corr$^{local}$, with the benchmark signals. For two signal series X and Y, it is defined as:

$$\text{Corr}^{local}(X, Y, m) = \frac{\sum_{i=m-M}^{m+M}(X_i - \bar{X}_{m-M}^{m+M})(Y_i - \bar{Y}_{m-M}^{m+M})}{(2M+1)\sigma_{X_{m-M}^{m+M}}\sigma_{Y_{m-M}^{m+M}}}, M < m \leq N - M,, \qquad (5)$$

where $\bar{X}_{m-M}^{m+M}$ and $\sigma_{X_{m-M}^{m+M}}$ represent the mean value and standard deviation of X from m-M to m+M, respectively. N is the length of X and Y. In this study, we use M=30.

One main purpose of extracting the respiratory signal is for the reconstruction of 4D-CBCT images. Hence, a study is necessary to investigate how the respiratory signals extracted by different methods may affect the quality of the reconstructed 4D-CBCT images. For this purpose, a typical patient case is selected where different methods yield different reconstruction results. The projections at the end of exhale (EE) are sorted out by respiratory signals extracted using different methods and then used to reconstruct a CBCT image with the FDK algorithm (Feldkamp *et al.*, 1984). A reference average image is also reconstructed using all the projections.

## 3. Results

### 3.1 Extracted breathing signals for scans with normal dose

Two sets of same ROIs were used for all methods: $ROI_1$ is the lower 1/3 of each projection image and $ROI_2$ is the whole projection image. Figure 5 shows the typical breathing signals extracted by various methods versus the RPM signals and the benchmark signals for patients 3 and 2. It can be seen that all methods work well for patient 3 where the diaphragm is included in the CBCT projection images. For patient 2, however, the diaphragm is not included in the projection images and thus some methods such as AS ($ROI_1$ and $ROI_2$) and FT-p ($ROI_1$) do not work well.

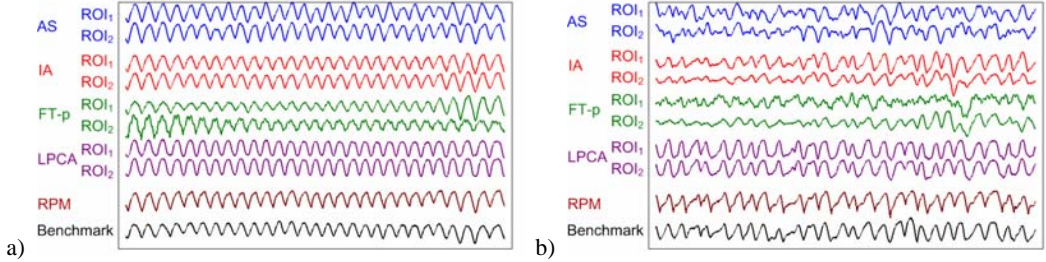

**Figure 5.** The breathing signals extracted by various methods (AS, IA, FT-p, and LPCA) are compared with the RPM signals and the benchmark signals for two typical example cases: a) patient 3 with diaphragm in the projection images and b) patient 2 without diaphragm in the projection images.

Figure 6 shows a quantitative evaluation in terms of $Corr^{local}$ for patients 1, 3 and 6 whose diaphragms were included in the CBCT projection images. We can see that 1) the AS method works well for patients 1 and 3 but not for the second half data of patient 6; 2) the IA method works well for patients 1 and 3 but not that well for patient 6; 3) the FT-p method works well for patient 1 with $ROI_2$ and patients 3 and 6 with $ROI_1$, but not for patient 1 with $ROI_1$ and patients 3 and 6 with $ROI_2$, indicating that this method is sensitive to the selection of ROI; 4) the proposed LPCA method works well for all three patients and both ROIs, showing the robustness of this method; and 5) the RPM signal correlates with the benchmark signal very well for patient 3 but less well for patient 1, indicating the existence of possible phase shift between internal and external signals.

Figure 7 presents the same evaluation for patient 2, 4 and 5 whose diaphragms are not included in the CBCT projection images. Here we can observe that 1) the AS method does not work well for any patients; 2) the IA method more or less works for patients 2 and 5 but not for patient 4; 3) the FT-p method works to certain degree for patient 2 with $ROI_2$ but not for other cases; 4) the LPCA method approximately works for almost all cases except for the second half of gantry rotation for patient 5 with $ROI_2$, where the anatomical breathing patterns in the AS image diminish due to the half-fan scanning mode; and 5) the RPM signal approximately correlates with the benchmark signal for patient 2 but does not correlate for patient 1, indicating that for some patients significant phase shift may exist between internal and external signals.

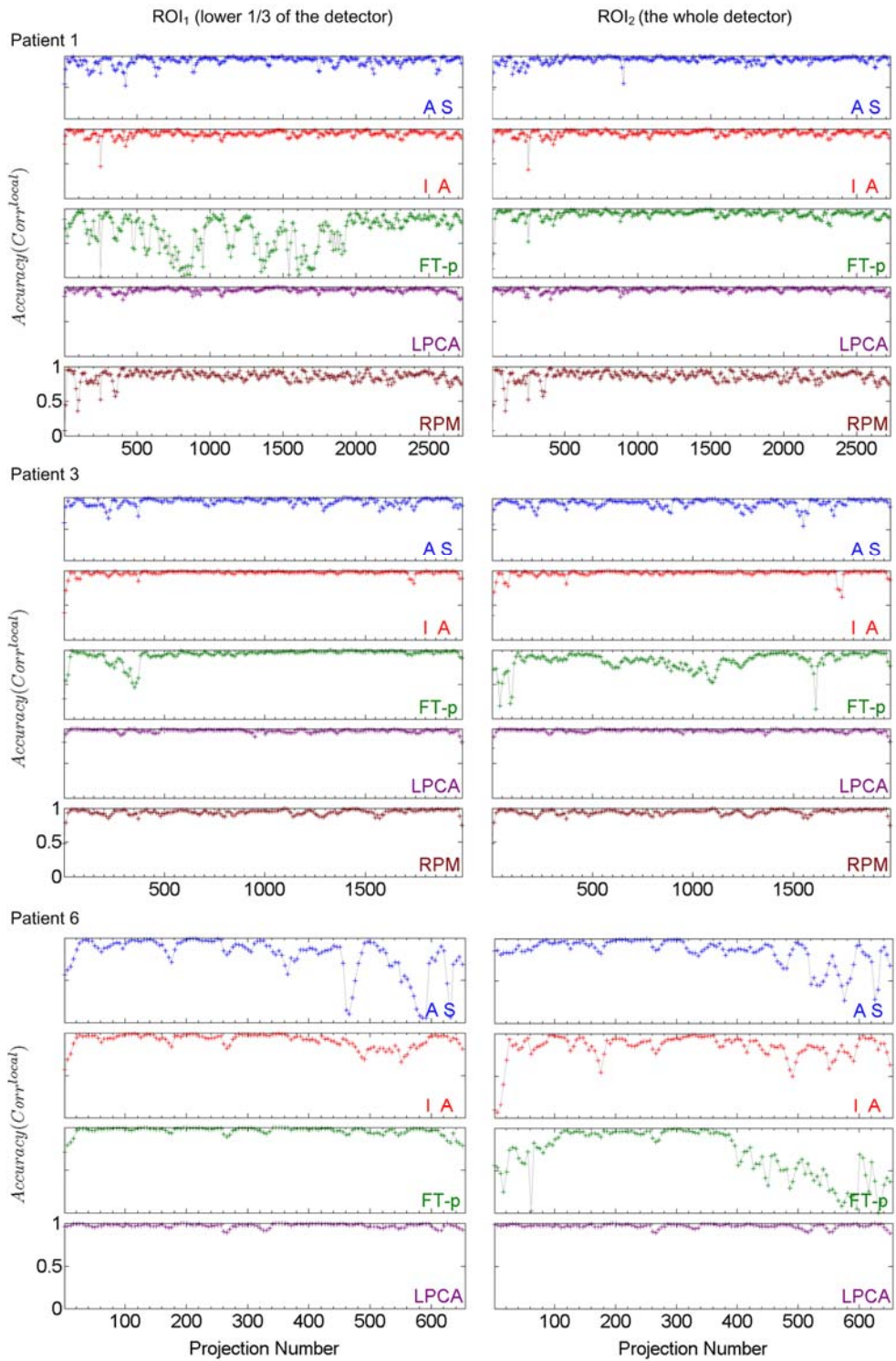

**Figure 6.** Local correlation coefficients (Corr$^{local}$) between the breathing signals extracted by various methods (AS, IA, FT-p, LPCA, and RPM) and the benchmark signals for patients 1, 3 and 6 (with diaphragm in the projection images) for two different ROI's (ROI$_1$ - left column and ROI$_2$ – right column). For every sub-plot the y axis ranges from 0 to 1.

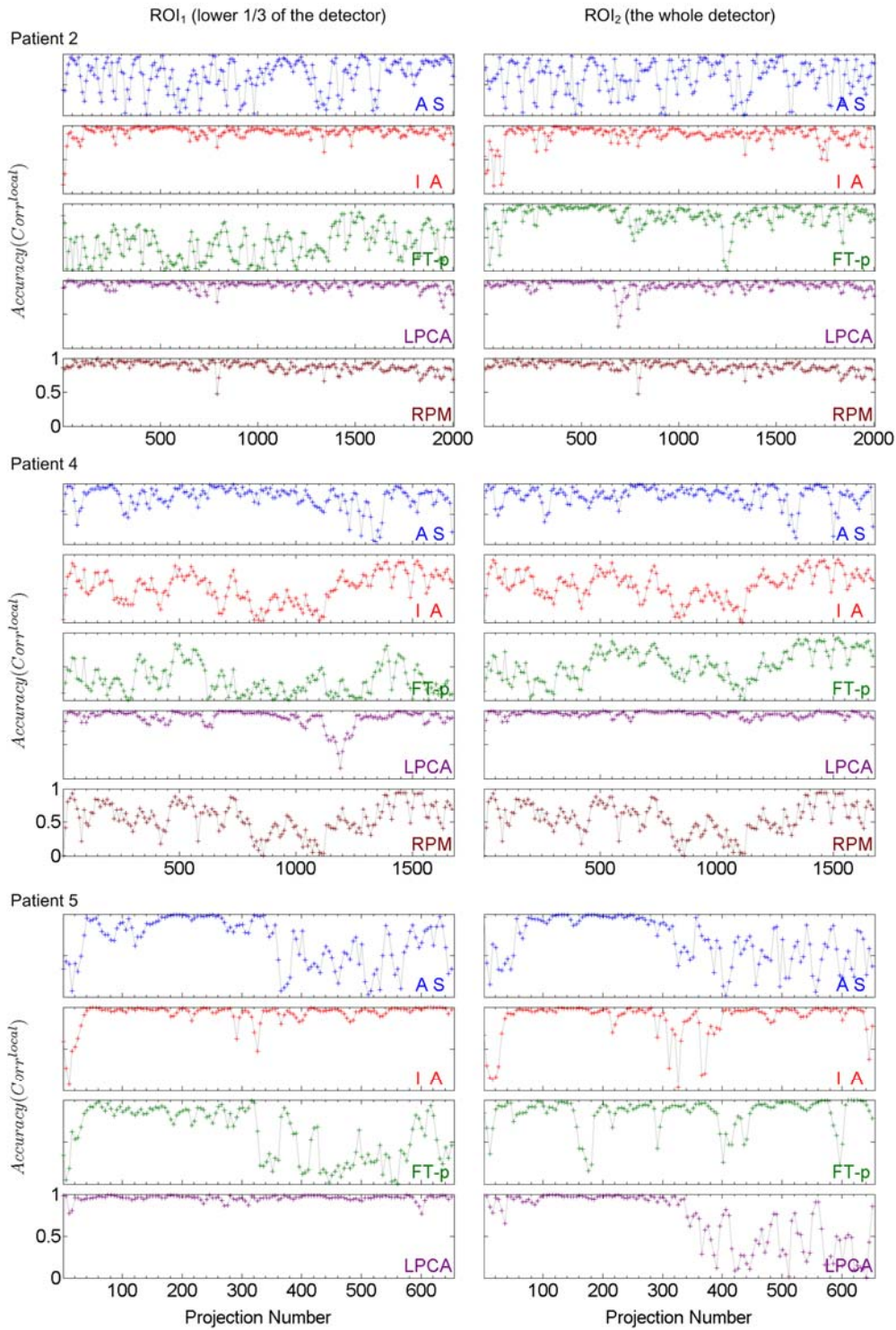

**Figure 7.** Local correlation coefficients (Corr^local) between the breathing signals extracted by various methods (AS, IA, FT-p, LPCA, and RPM) and the benchmark signals for patients 2, 4 and 5 (without diaphragm in the projection images) for two different ROI's (ROI$_1$ - left column and ROI$_2$ – right column). For every sub-plot the y axis ranges from 0 to 1.

*3.2 Extracted breathing signals for scans with low dose*

For the low-dose data (patients 7 and 8), qualitative visual inspection is performed by displaying all extracted signals side by side with the corresponding AS images (figure 8). Based on the visual inspection, we found that, for the gantry angles with extremely weak signals as indicated by the yellow boxes in figure 8, 1) the AS method does not work for either case; 2) the FT-p method works well for patient 8 with $ROI_2$, less well for patient 8 with $ROI_1$, and does not work for patient 7; and 3) the IA and LPCA methods work for all cases except for patient 7 with $ROI_1$.

It can also be seen that for patient 7, a better result is obtained with $ROI_2$ than with $ROI_1$ by using either IA or LPCA. A similar observation can be made for patient 8 too using FT-p. This finding indicates that a larger ROI may be preferred for gantry angles with extremely weak signals.

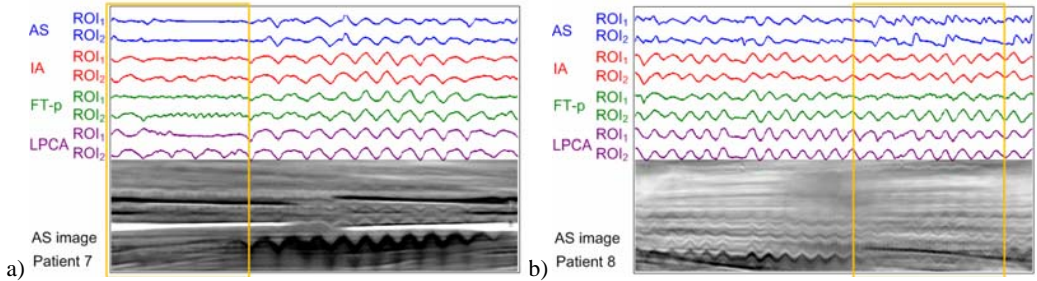

**Figure 8.** The breathing signals extracted by various methods (AS, IA, FT-p, and LPCA) are displayed together with the corresponding AS images for visual inspection for low-dose CBCT scans (patients 7 and 8). Yellow boxes highlight the gantry angles with extremely weak signals in the AS images.

### 3.3 Effect of different ROI sizes

We evaluated the effect of different ROI sizes on the performance of the AS, IA, FT-p and LPCA methods. The results are shown in figure 9. In each sub-figure, the y-axis represents the averaged local correlation coefficients ($\overline{Corr^{local}}$) between the extracted signals using various methods and the benchmark signals. The x-axis represents the ROI size normalized to the whole detector along the SI direction. The ROI covers the whole detector along the transverse direction and the bottom part of the detector along the SI direction. Yellow vertical lines indicate $ROI_1$ and $ROI_2$ used in previous sections.

We can see that: 1) the AS and IA methods are less dependent on the ROI size but for some cases (patients 2, 5 for AS and patient 4 for IA) these two methods do not perform well; 2) the FT-p method is sensitive to the selection of ROI especially for patients 2, 4, and 5; 3) the proposed LPCA method performs extremely well for all ROI sizes and for all patients except for patient 4 with small ROI sizes and patient 5 with very small and very large ROI sizes.

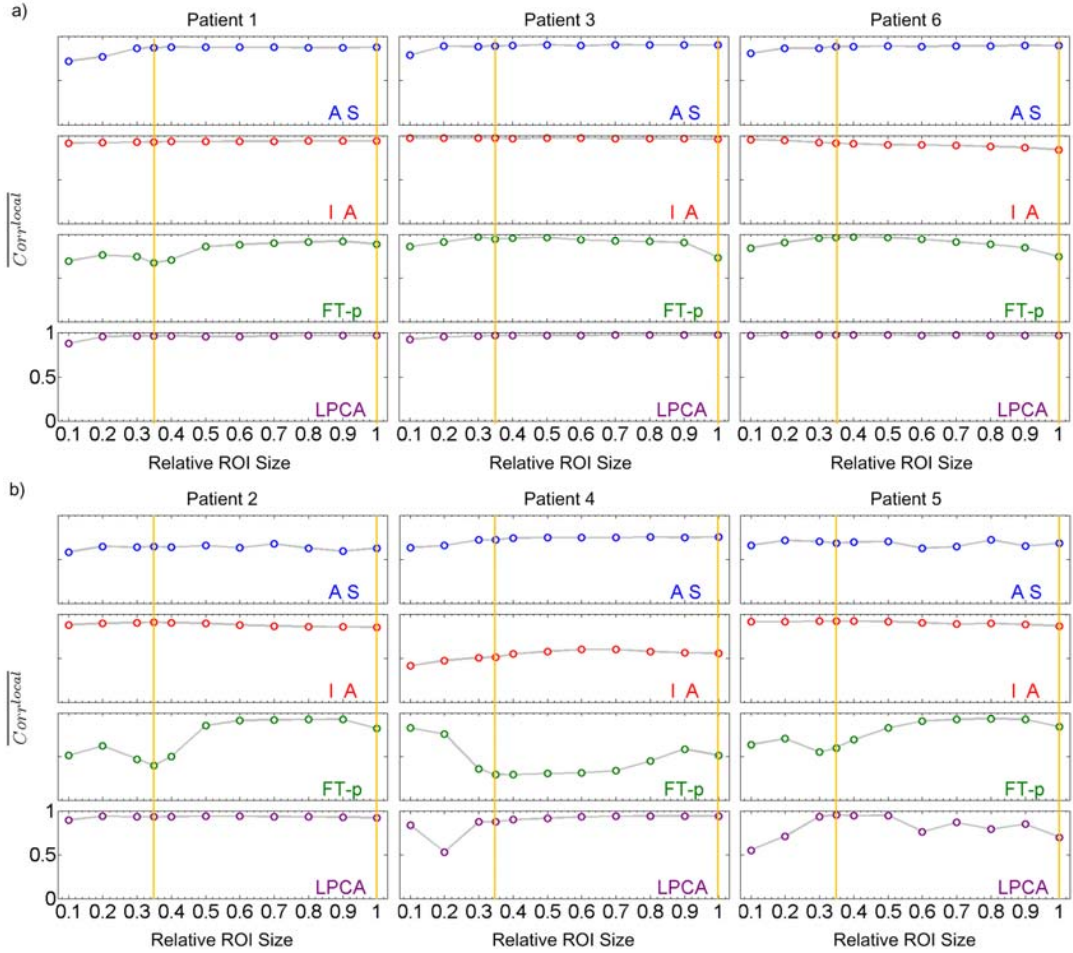

**Figure 9.** The averaged local correlation coefficients ($\overline{\text{Corr}^{local}}$) between the signals extracted using various methods from different ROIs and the benchmark signals. The yellow vertical lines denote the ROI$_1$ and ROI$_2$ used in previous sections. a) Patients 1, 3 and 6 with diaphragms in the projection images; b) Patients 2, 4 and 5 without diaphragms in the projection images. For every sub-plot the y axis ranges from 0 to 1.

### 3.4 Effect on the quality of 4D-CBCT images

Patient 4 is used in this study. The average image and the images at the EE phase reconstructed using projections binned by different respiratory signals are shown in figure 10. For each row, a base line is plotted to help comparing the locations of objects. In the selected ROIs, three highlighted objects A, B and C are indicated by arrows.

It can be seen that the different methods give different results, and we notice that: 1) The LPCA method generates the most accurate locations of the internal structures along the SI direction as indicated by the highest SI location of the internal structures (e.g. the object C) in the reconstructed images. Among all EE images reconstructed using different respiratory signals, the one with internal structures at their most superior positions should be the most accurate. We also measure the SI shift of the structure C between the results of the LPCA and IA method, which is 4.4 mm, an in-negligible error. In fact, we observe that the breathing signals derived from other methods have phase shifts at some breathing cycles from the ground truth, which contributes to the location error seen here. 2) The LPCA method is most effective in terms removing motion caused blurring artifacts and offers the clearest images (see the structure A). 3) Overall, the performances of these method in terms of the resulting 4D-CBCT image quality are consistent with the $\overline{\mathrm{Corr}^{\mathrm{local}}}$ results (figure 7). Specifically, AS method is better than the other three existing methods in this case.

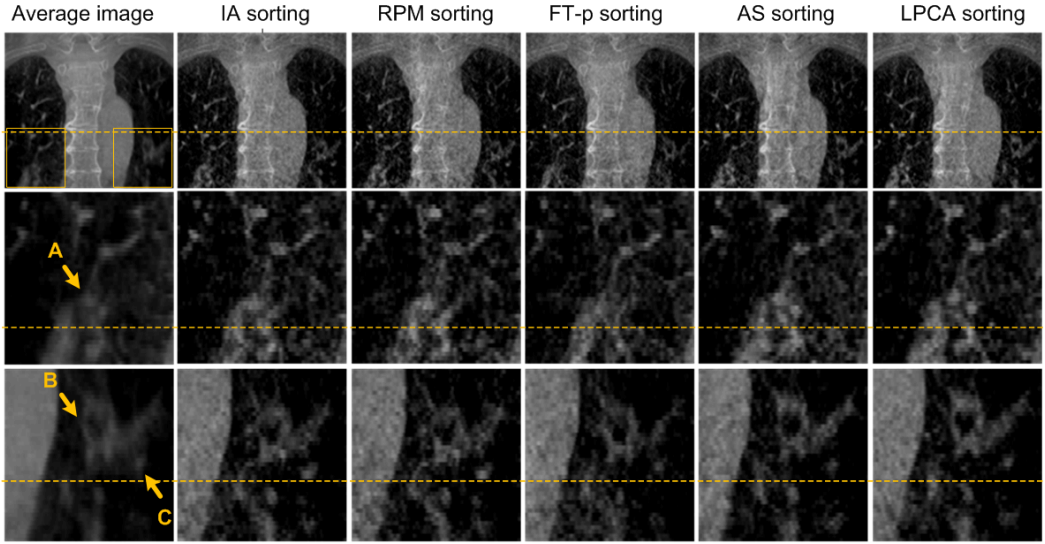

**Figure 10.** Reconstructed 4D-CBCT image at the end expiration phase for patient 4. Rows from top to bottom: the coronal view of the CBCT images, zoom-in images corresponding to the ROI in right lung in the first image of the top row, and zoom-in images corresponding to the ROI in the left lung. Columns from left to right: the averaging image reconstructed with all projections, the images reconstructed using projections sorted by IA, RPM, FT-p, AS and LPCA methods, respectively. For each row, a base line is plotted to help comparing the locations of objects. In the selected ROIs, three investigated objects are highlighted by arrows A, B and C.

### 3.5 LPCA: effect of the parameters and computational time

Figure 11 shows the influence of different sliding window width $W$ on the results of the proposed LPCA method. The y-axis represents $\overline{\mathrm{Corr}^{\mathrm{local}}}$ between the LPCA-extracted signals and the benchmark signals while the x-axis represents the sliding window width in the unit of the number of breathing cycles. We can see that once $W$ goes beyond 1 breathing cycle, the accuracy reaches an expected high level, consistent with the analysis for $W$ value selection in Section 2.2.2. The high accuracy is maintained within a large range of $W$ for patients 1-4 (4-minute scan). For patient 5 and 6 (1-minute scan), the range is relatively small. There are two reasons for this fact. First, $W$ should be in principle small enough, such that within this range the torus structure can be well approximated by a linear space. For these two patient cases where gantry rotates relatively fast, the window width of a few breathing cycle correspond to a large scan angles, where the curved geometry of the torus start to appear within this angular range. Second, these two patients are scanned in the half-fan mode. Hence the projection of the patient body changes relatively more dramatically as the gantry rotates, making the PCA less effectively. Based on all of these cases, we found a range of $W$ between 1 and 3 breathing cycles is feasible for all cases.

The computational time of LPCA code is tested under MatLab 2010b on a desktop PC with 3.17GHz CPU and 8 GB RAM. The average time for extracting the breathing signal from one projection is 72.8 ms. This value is expected to be smaller through efficiency optimization.

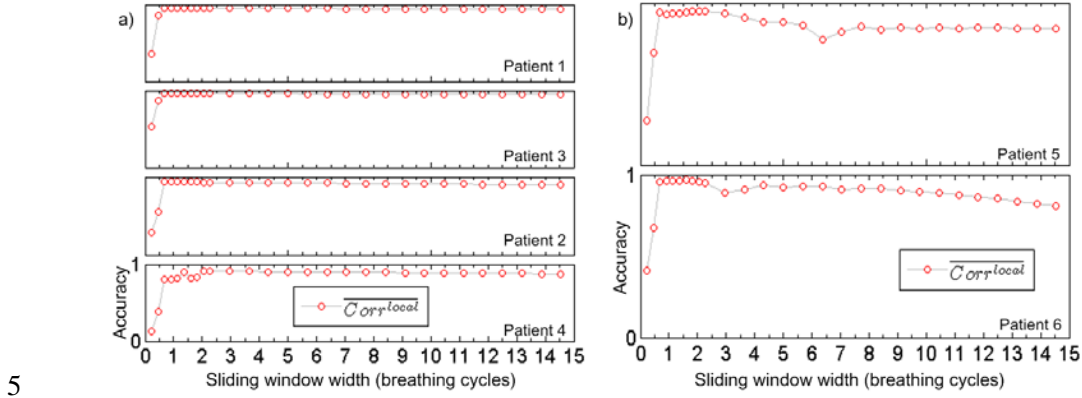

**Figure 11.** The averaged local correlation coefficients ($\overline{Corr^{local}}$) between the LPCA-extracted signals and the benchmark signals for different the sliding window width $W$. The vertical range of each sub-plot is [0, 1]. $W$ is in the unit of the number of breathing cycles. (a). 4-minute scan data (patients 1-4). (b). 1-minute scan data (patient 5 and 6).

## 4. Discussion

From the comparisons (figures 6-9), we can see that the IA method has stable yet slightly less accurate performances comparing to LPCA. While the AS method yields a signal with clear physical meaning, namely, the motion of certain anatomical features in lung along the SI direction, its effectiveness relies on the distinctness of the moving features such as diaphragm. This observation is consistent with that of (Kavanagh *et al.*, 2009). The performance of FT-p method depends on the ROI selection, which reminds us that the segmentation of the lung region in each projection image may be crucial for the success of this method (Vergalasova *et al.*, 2012).

The proposed LPCA method is validated as a promising projection-based method with the best overall performance among all the tested cases (figures 6-10). Compared with the RPM signal available for patients 1-4, LPCA shows better or at least similar correlation with the internal anatomical features (figure 6, 7). Particularly for patient 4, LPCA shows impressive improvement on the 4DCBCT images reconstructed using the derived breathing signals (figure 10). Besides the normal dose data, we have also tested two low dose cases. The excessive noise in such cases does not introduce apparent compromise to the results (figure 8). The underlying reasons might be: 1) A 4×4 pixel binning was used in each projection to suppress the noise; 2) To get the AS image, a horizontal average in projection images is needed, and this process is able to suppress noise too.

While LPCA could achieve impressive overall performance in terms of breathing signal extraction, it did not work well in certain circumstances, e.g. patient 5 with ROI$_2$ (figure 7) and patient 7 with $ROI_1$ (figure 8). A potential solution is to combine the LPCA method with other methods to improve the robustness. This will be investigated in our future work.

5       It is difficult to extract breathing signals from CBCT projection images in half-fan scan mode (patients 5, 7, and 8, figure 4). This issue is more serious under low-dose scan protocol, where the low exposure could not provide enough contrast for lateral views, causing unclear wave patterns in that part of the AS images (patients 7 and 8, figures 4 and 8). This poses a great challenge to all the methods. In such cases, a larger ROI seems preferred for all the effective methods (figure 8).

10      In additional to 4D-CBCT, extracting respiratory signal is also useful for real-time tumor tracking. In such a non-retrospective scenario, LPCA method needs a pre-acquisition longer than one breathing cycle (Section 2.2.2 and figure 11) to obtain part of the AS image and hence to facilitate the required image processing. Comparing with the 200-ms systematic delay tolerance for real time tumor tracking (Murphy *et al.*, 2002;
15      Sharp *et al.*, 2004; Vedam *et al.*, 2004; Keall *et al.*, 2006), the un-optimized LPCA computational time (72.8 ms) is of a reasonable level. A further optimization of the code will be conducted in our future work to make it feasible for tracking tumor in real time based on this method.

### 5. Conclusions

20      The proposed LPCA method is validated as a promising technique in extracting respiratory signals, with a best overall performance for all eight tested patient cases. Each of the existing methods has its own applicability. AS method relies on distinguished moving anatomical features and hence favors the presence of diaphragm in the projection images. The performance of FT-p method is sensitive to the ROI selection and the
25      segmentation of lung region in each projection image may be crucial for its success. IA method has a stable performance, slightly less accurate than LPCA. For the cases where the anatomy movement information is extremely weak or inconsistent, a larger ROI is preferable. As for the clinical applications, it may need to incorporate multiple methods together to improve the robustness and accuracy. We would like to mention that the
30      computer code for the LPCA method is available upon request for the research community.

### Acknowledgements

This work is supported in part by NIH (1R01CA154747-01) and Varian Medical Systems through a Master Research Agreement. The work of W. Yin is supported in part by NSF
35      grants DMS-0748839 and ECCS-1028790, ONR grant N00014-08-1-1101.